\documentclass[francais]{gretsi}

\usepackage[utf8]{inputenc}

\usepackage[T1]{fontenc}
\usepackage[french,english]{babel}

\usepackage{amsmath, amssymb, amsfonts}

\usepackage{algorithm}
\usepackage{algpseudocode}

\usepackage{cite}

\begin{document}


\titre{Implémentation Efficiente de Fonctions de Convolution sur FPGA à l’Aide de Blocs Paramétrables et d'Approximations Polynomiales}

\auteurs{
  \auteur{Philippe}{Magalhães}{philippe.magalhaes@univ-st-etienne.fr}{1}
  \auteur{Virginie}{Fresse}{virginie.fresse@univ-st-etienne.fr}{1}
  \auteur{Benoît}{Suffran}{benoit.suffran@st.com}{2}
  \auteur{Olivier}{Alata}{olivier.alata@univ-st-etienne.fr}{1}
}

\affils{
  \affil{1}{Laboratoire Hubert Curien, UMR 5516, CNRS, IOGS, Univ. J. Monnet,
        18 Rue Prof. Benoît Lauras,
        42000 Saint-Etienne, France
  }
  \affil{2}{ST Microelectronics,
        12 Rue Jules Horowitz,
        38019 Grenoble, France
  }
}


\resume{La mise en œuvre de réseaux de neurones convolutifs (CNNs) sur des \textit{field-programmable gate arrays} (FPGAs) représente une solution alternative aux GPU, en raison d'une latence réduite, une meilleure efficacité énergétique et une flexibilité accrue. Cependant, elle reste complexe en raison des compétences matérielles requises et des longues phases de synthèse, placement et routage, qui allongent les cycles de conception et limitent l’exploration rapide des configurations, compliquant l’optimisation sous fortes contraintes. Cet article propose une bibliothèque de Blocs de convolution configurables, visant à optimiser l’implémentation sur FPGA et à s’adapter aux ressources disponibles. Il présente également un cadre méthodologique pour développer des modèles mathématiques capables de prédire l’utilisation des ressources FPGA. L’approche est validée par une analyse de corrélation entre les paramètres suivie de mesures d’erreur. Les résultats montrent que les blocs conçus permettent l’adaptation des convolutions aux contraintes matérielles, et que les modèles prédisent avec précision la consommation des ressources, constituant un outil utile pour le choix d’architectures FPGA et le déploiement optimisé des CNNs.}

\selectlanguage{english}
\abstract{Implementing convolutional neural networks (CNNs) on field-programmable gate arrays (FPGAs) has emerged as a promising alternative to GPUs, offering lower latency, greater power efficiency and greater flexibility. However, this development remains complex due to the hardware knowledge required and the long synthesis, placement and routing stages, which slow down design cycles and prevent rapid exploration of network configurations, making resource optimisation under severe constraints particularly challenging. This paper proposes a library of configurable convolution Blocks designed to optimize FPGA implementation and adapt to available resources. It also presents a methodological framework for developing mathematical models that predict FPGA resources utilization. The approach is validated by analyzing the correlation between the parameters, followed by error metrics. The results show that the designed blocks enable adaptation of convolution layers to hardware constraints, and that the models accurately predict resource consumption, providing a useful tool for FPGA selection and optimized CNN deployment.}

\maketitle

\section{Introduction}

Les CNN sont largement utilisés dans des domaines tels que la vision artificielle, la reconnaissance des formes et le traitement du signal. Ils permettent d'améliorer de manière significative la précision des résultats. Bien que traditionnellement implémentés sur GPU, les FPGA constituent une alternative prometteuse grâce à leurs avantages en matière de débit, d'efficacité énergétique, de réactivité et de flexibilité, des qualités essentielles dans les systèmes embarqués et pour le temps réel. \cite{shao2024}\cite{ye2024}\cite{xu2024}\cite{AlAli2020}.

Plusieurs travaux se concentrent sur l'optimisation de l'implémentation des CNN sur FPGA. Xu et al. \cite{xu2024} proposent un modèle d'exploration de l'espace de conception (DSE) pour estimer l'utilisation des ressources, ce qui permet une reconfiguration dynamique des couches afin de maximiser l'accès aux ressources FPGA requises pour les opérations. Par ailleurs, Li et al. \cite{li2024} exploitent le DSE pour localiser les régions de calcul critiques et définir des stratégies de partitionnement optimales pour la reconfiguration dynamique sur FPGA, dans le but de réduire la consommation d'énergie.

Shi et al. \cite{shi2023} introduisent la reconfiguration dynamique dans des régions du FPGA afin de résoudre les problèmes de ressources matérielles insuffisant. Shao et al.\cite{shao2024} emploient des quantifications sur entiers 8 bits (INT8) et 16 bits (INT16), combinées à une matrice systolique, afin d’atteindre une meilleure efficacité énergétique. Tandis que Fuketa et al. \cite{fuketa2024} a utilisé la quantification INT8 sans DSPs, atteignant une forte utilisation des LUTs et BRAM sur ZCU104, avec une perte de précision de 5\%. Hamdan et al.\cite{hamdan2017} ont développé un outil de génération VHDL pour optimiser les implémentations CNN sur FPGA, et Liu et al. \cite{liu2024} explorent des techniques d’optimisation des boucles pour réduire la latence et améliorer l’usage de la mémoire embarquée. La Table \ref{tab:utilisation_ressources} résume les modèles et les architectures utilisés, ainsi que la consommation de ressources rapportée dans certains des travaux précédemment cités.

Le déploiement des CNNs sur FPGA reste complexe, malgré la disponibilité d’outils capables de générer automatiquement du code ou d’estimer certaines métriques. En pratique, ces solutions peuvent parfois ne pas garantir une allocation optimale des ressources ni une adéquation idéale entre le réseau et le matériel, entraînant souvent une sous-utilisation ou une saturation du FPGA. Le processus exige aussi une expertise matérielle et des phases longues de synthèse, de placement et de routage, freinant l’exploration de l’espace de conception.

Pour y répondre, une bibliothèque de blocs de convolution réutilisables et un cadre méthodologique sont proposés pour estimer la consommation de ressources à partir du nombre de bits des données et des coefficients. Cette approche peut orienter le choix des blocs de convolution ou de la plateforme FPGA, faciliter l’adaptation des couches aux contraintes matérielles, réduire les temps de développement et les itérations de synthèse inutiles.

\begin{table}[h]
    \centering
    \caption{Utilisation des ressources pour différentes implémentations de CNN sur plateformes FPGA.}
    \label{tab:utilisation_ressources}
    \begin{tabular}{p{0.5cm}|p{1.9cm}|p{1.4cm}|p{0.8cm}|p{0.8cm}|p{0.7cm}}
        \hline
        \small{\textbf{Réf.}} & \small{\textbf{Réseau}} & \small{\textbf{Plateforme}} &\small{\textbf{LUT (\%)}}&\small{\textbf{FF (\%)}}& \small{\textbf{DSP (\%)}} \\
        \hline
        \cite{shi2023}   & \small{YOLOv2-Tiny}        & KV260     & 99,4  & 100   & 100  \\
        \cite{shao2024}   & \small{YOLOv3-Tiny(INT8)}  & VC709     & 39,0  & 16,10 & 14,28\\
        \cite{shao2024}  & \small{YOLOv3-Tiny(INT16)} & VC709     & 51,73 & 20,00 & 28,56\\
        \cite{fuketa2024} & \small{RLDA}               & ZCU104    & 88,2  & 33,4  & 0,0  \\
        \cite{hamdan2017} & \small{LeNet}              & Virtex-7  & 61,05 & 27,02 & 2,08 \\
        \cite{hamdan2017} & \small{AlexNet}            & Virtex-7  & 66,35 & 31,14 & 57,5 \\
        \cite{liu2024}    & \small{VGG-16}             & ZCU102    & 51,38 & 16,64 & 20,31\\
        \cite{liu2024}    & \small{VGG-16}             & ZCU111    & 73,88 & 18,66 & 47,94\\
        \hline
    \end{tabular}
\end{table}

\section{FPGA}

Les FPGA offrent une plateforme versatile et optimisée pour les opérations logiques et arithmétiques. Leur structure repose sur des blocs logiques configurables, qui comprennent des LUTs pour la logique combinatoire (LLUTs) et la mémoire distribuée (MLUTs), des flip-flops (FF) pour le stockage et la synchronisation, ainsi que des multiplexeurs qui permettent une interconnexion flexible.

Les \textit{Carry Chains} (CChains) facilitent les calculs arithmétiques, notamment les additions, en transférant rapidement les retenues d’un bloc logique à l’autre, ce qui accélère le traitement sans utiliser de ressources complexes.

Les ressources mémoire comprennent différents types de RAM, utilisés pour le stockage centralisé ou distribué des données, selon les besoins de capacité et de performance.

Enfin, le \textit{Digital Signal Processor} (DSP) permet des calculs arithmétiques complexes, tels que multiplication et accumulation, avec flexibilité en précision et fréquence élevée.

Chaque FPGA ayant une architecture et une disponibilité de ressources qui lui sont propres, une bonne compréhension de ces spécificités est essentielle pour adapter efficacement l’implémentation aux capacités matérielles.

\section{Méthodologie}
La méthodologie comprend quatre étapes : 1) le développement de blocs paramétrables pour explorer différentes stratégies d’implémentation selon les contraintes matérielles ; 2) la collecte des données de consommation via synthèse ; 3) l'analyse de corrélation pour identifier les paramètres influents et orienter le choix du type de modèle ; 4) la construction de modèles mathématiques permettant d’estimer l’utilisation des ressources FPGA à partir des configurations des blocs.

\subsection{Blocs de convolution}

Afin de répondre à l’hétérogénéité  des plateformes FPGA et d’explorer différentes stratégies d’implémentation, quatre blocs de convolution paramétrables ont été développés en VHDL. Ils utilisent l’arithmétique en virgule fixe, avec un chargement série et stockage local des coefficients du noyau 3×3 pour optimiser la mémoire, et un chargement parallèle des données pour maximiser le débit. Ces blocs se distinguent par leur utilisation des ressources FPGA, notamment le bloc $Conv_1$, qui est le seul à exploiter des CChains sans recourir aux blocs DSP. La Table~\ref{tab:convolution_ips} résume leurs caractéristiques.

\begin{table}[h]
\centering
\caption{Caractéristiques des blocs de convolution.}
\label{tab:convolution_ips}
\begin{tabular}{|p{0.85cm}|p{1.2cm}|p{1.5cm}|p{3.6cm}|}
\hline
\textbf{Bloc} & \textbf{Usage}   & \textbf{Usage de}   & \textbf{Caractéristiques}    \\ 
            & \textbf{du DSP}  & \textbf{la logique} &  \textbf{principales}        \\   \hline
$Conv_1$    & Aucun            & Haut                & Logique et CChains ;   \\ 
            &                  &                     & une convolution par cycle.   \\ \hline
$Conv_2$    & 1 DSP            & Faible              & Logique réduite ;            \\ 
            &                  &                     & une convolution par cycle.   \\ \hline
$Conv_3$    & 1 DSP            & Modéré                & 2 convolutions parallèles ;  \\ 
            &                  &                     & Opérandes jusqu'à 8 bits.     \\ \hline
$Conv_4$    & 2 DSPs           & Modéré              & 2 convolutions parallèles,  \\ 
            &                  &                     & une par DSP.       \\ \hline
\end{tabular}
\end{table}

\subsection{Collecte de données de synthèse}
Pour cette étape, 196 configurations de fonction de convolution ont été utilisées, générées en faisant varier la taille des données et les coefficients de 3 à 16 bits. La synthèse a été réalisée avec Vivado 2024.2 sur Zynq UltraScale+ ZCU104, et les mesures de consommation de ressources, incluant les LLUT, MLUT, FF, CChains et les DSP, ont été enregistrées pour chaque configuration et type de bloc.

\subsection{Analyse des données}

L’analyse de corrélation de Pearson a permis d’évaluer l’impact des tailles de données et de coefficients sur la consommation des ressources FPGA. Les résultats de la Table~\ref{tab:pearson_zcu104} montrent une corrélation élevée pour la plupart des blocs, avec des valeurs supérieures à 0,65 pour les LLUTs et MLUTs, indiquant une relation linéaire significative. Les flip-flops présentent également une forte corrélation avec la taille des coefficients. À l’inverse, $Conv_3$ présente une corrélation nulle avec la taille des données et modérée avec les coefficients pour LLUTs et MLUTs, suggérant une relation non linéaire, possiblement liée au parallélisme sur un seul DSP.

\begin{table}[h]
\caption{Corrélation de Pearson.}
\label{tab:pearson_zcu104}
\begin{tabular}{p{0.8cm}p{1.4cm}p{1.4cm}p{0.8cm}p{0.8cm}p{0.8cm}}
\toprule
\textbf{$Conv_1$}  & \textbf{Taille des}  & \textbf{Taille des} & \textbf{LLUT} & \textbf{MLUT} & \textbf{CChain} \\ 
                   & \textbf{données} &  \textbf{coeffs} &                &                &                   \\ 
\midrule
LLUT    & 0,668 & 0,672 &  &  &  \\  
MLUT    & 0,668 & 0,672 & 1,000 &  &  \\  
CChain  & 0,672 & 0,672 & 0,981 & 0,981 &  \\  
FF      & 0,680 & 0,733 & 0,947 & 0,947 & 0,950 \\ 
\bottomrule
\hline
\end{tabular}

\vspace{0.1cm}

\begin{tabular}{p{1cm}p{1.5cm}p{1.5cm}p{1cm}p{1cm}}
\toprule
\textbf{$Conv_2$}  & \textbf{Taille des données} & \textbf{Taille des coeffs} & \textbf{LLUT} & \textbf{MLUT} \\ 
\midrule
LLUT    & 0,658 & 0,713 &  &  \\ 
MLUT    & 0,658 & 0,713 & 1,000 &  \\ 
FF      & 0,000 & 0,997 & 0,718 & 0,718 \\ 
\bottomrule
\hline
\end{tabular}

\vspace{0.1cm}

\begin{tabular}{p{1cm}p{1.5cm}p{1.5cm}p{1cm}p{1cm}}
\toprule
\textbf{$Conv_3$}  & \textbf{Taille des données} & \textbf{Taille des coeffs} & \textbf{LLUT} & \textbf{MLUT} \\ 
\midrule
LLUT    & 0,000 & 0,497 &  &  \\ 
MLUT    & 0,000 & 0,497 & 1,000 &  \\ 
FF      & 0,000 & 0,996 & 0,495 & 0,495 \\ 
\bottomrule
\hline
\end{tabular}

\vspace{0.1cm}

\begin{tabular}{p{1cm}p{1.5cm}p{1.5cm}p{1cm}p{1cm}}
\toprule
\textbf{$Conv_4$}  & \textbf{Taille des données} & \textbf{Taille des coeffs} & \textbf{LLUTs} & \textbf{MLUTs} \\ 
\midrule
LLUT    & 0,691 & 0,714 &  &  \\ 
MLUT    & 0,691 & 0,714 & 1,000 &  \\  
FF      & 0,000 & 0,997 & 0,715 & 0,715 \\ 
\bottomrule

\end{tabular}
\end{table}

Ces résultats permettent non seulement d’identifier les variables d’entrée les plus influentes sur la consommation de ressources, mais aussi d’anticiper le type de modèle adapté à chaque bloc. Une corrélation élevée justifie l’usage de modèles polynomiaux simples, tandis qu’une corrélation faible nécessite des approches non linéaires ou segmentées.

\subsection{Construction des modèles}

Des modèles polynomiaux ont été construits pour chaque configuration de bloc afin d’estimer la consommation des ressources FPGA. Les LLUTs ont été choisies comme variable cible en raison de leur rôle central dans la logique combinatoire, de leur forte corrélation avec d’autres ressources, et de leur capacité à refléter l’utilisation globale sans perte de généralité. Le type de modèle a été défini selon les corrélations observées : des régressions polynomiales \cite{montgomery} ont été utilisées pour les blocs dont l’utilisation des ressources est linéairement corrélée avec la taille des entrées, et une régression segmentée pour $Conv_3$.

\begin{figure}[H]
    \centering
    \includegraphics[width=0.37\textwidth]{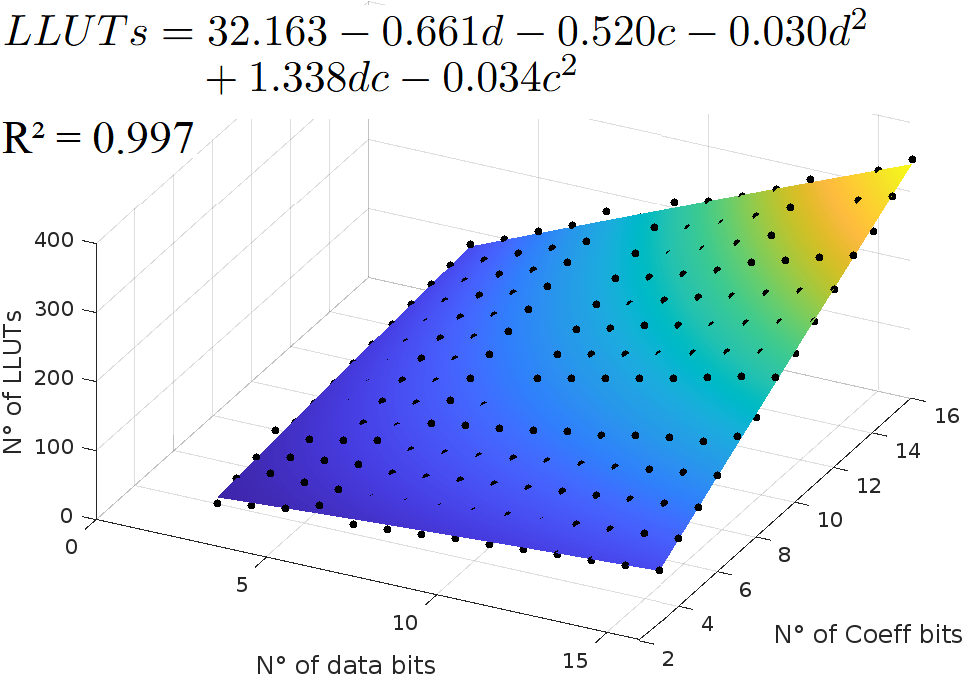}
    \caption{Consommation de LLUT - $Conv_1$.}
    \label{fig:conv1}
\end{figure}

\begin{figure}[H]
    \centering
    \includegraphics[width=0.37\textwidth]{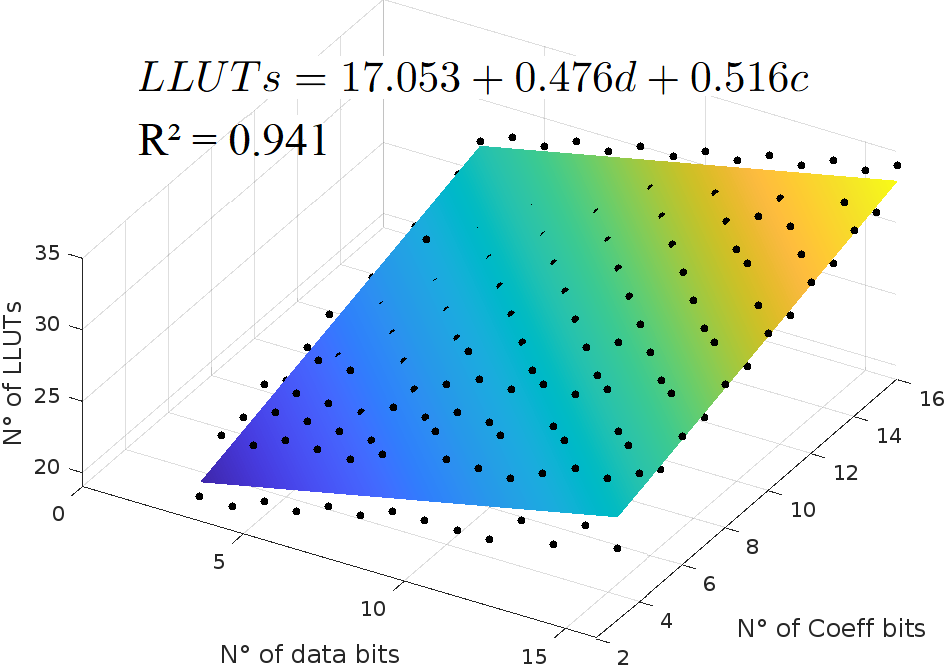}
    \caption{Consommation de LLUT - $Conv_2$.}
    \label{fig:conv2}
\end{figure}

\begin{figure}[H]
    \centering
    \includegraphics[width=0.37\textwidth]{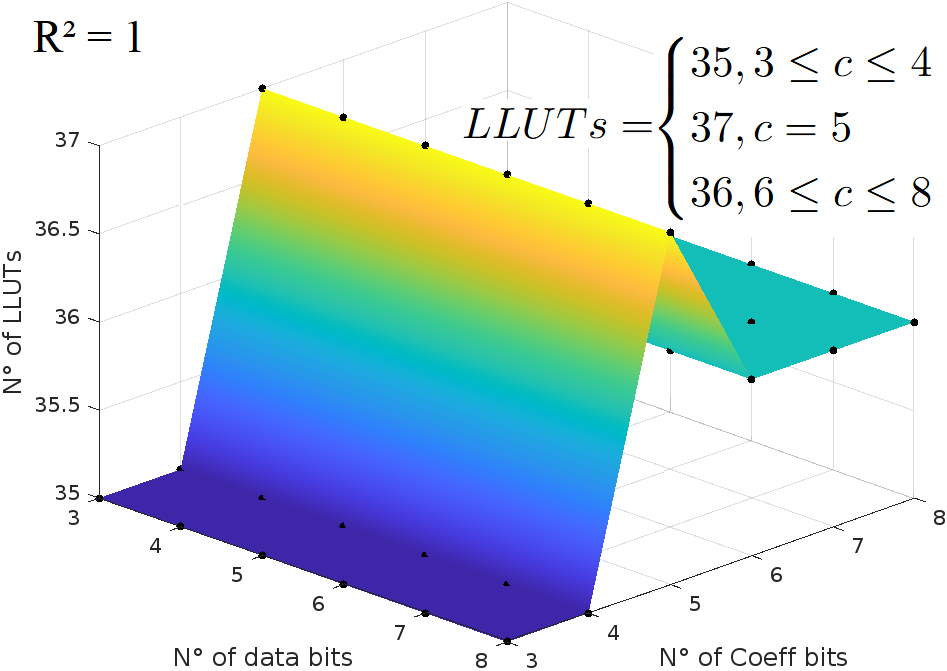}
    \caption{Consommation de LLUT - $Conv_3$.}
    \label{fig:conv3}
\end{figure}

Des polynômes de degré un à quatre ont été obtenu pour chaque ressource et chaque bloc, et les modèles avec un \( R^2 \) supérieur et proche de $0,9$ ont été retenus. L’Algorithm~\ref{alg:model_fitting} résume cette procédure, et les Figures~\ref{fig:conv1}, \ref{fig:conv2} et \ref{fig:conv3} présentent les nuages de points de consommation réelle de LLUTs, ainsi que les surfaces ajustées par les modèles pour $Conv_1$, $Conv_2$ et $Conv_3$. Le modèle pour $Conv_4$ est similaire à celui de $Conv_2$ et donné par l’équation $LLUTs = 20,886 + 1,004d + 1,037c$, avec $R^2 = 0,989$.

\begin{algorithm}[h]
\caption{Ajustement du modèle pour la prédiction des ressources FPGA.}
\label{alg:model_fitting}
\Pour{IP in \{$Conv_1, Conv_2, Conv_3, Conv_4$\}}{
    \Pour{FPGA=\{Plateformes\}}{
            \Pour{bit\_données $= 3 ... 16$}{
                \Pour{bit\_coeffs $= 3 ... 16$}{
                
                \small {
                    Synthétiser: (IP, FPGA, bit\_données, bit\_coeffs)\;
                                        
                    Mesurer et stocker l'utilisation des ressources;

                    }
                }
            }
        
        \Pour{\small{ressource=\{LLUT, MLUT, CChain, DSP, FF\}}}{
            
            $meilleur\_R^2 = 1$\;
            
            \textit{meilleur\_modèle}$=\texttt{NULL}$\;
            
            \Pour{degré=$1...4$}{
            
                \small{
                \textit{modèle = AjusterPolynôme (IP, ressource, degré)}\;
                }
                
                $R^2 = calculer R^2$ \textit{(modèle, IP, resource)}\;
                
                \Si{$ 0.9 \leq R^2 < meilleur\_R^2$}{
                
                    $meilleur\_R^2 = R^2$\;
                    
                    \textit{meilleur\_modèle = modèle}\;       
                } 
            }
            \small{
            \textit{s\_modèle = SupprimerInsignifiant (meilleur\_modèle)}\;
           
            $s\_R^2=calculerR^2$\textit{(s\_modèle, IP, ressource)}\;
            }
            
            \Si{$s\_R^2 \geq 0.9$}{
                \textit{meilleur\_modèle = s\_modèle}\;
            }
        }
    }
}
\end{algorithm}

\section{Validation}
\subsection{Mesures de performance}

Pour évaluer la précision et la fiabilité des modèles, quatre mesures de performance sont utilisées : l'erreur quadratique moyenne (EQM), l'erreur absolue moyenne (EAM), le coefficient de détermination ($R^2$) et l'erreur absolue moyenne en pourcentage (EAMP). Ces mesures sont essentielles pour garantir que les estimations de la consommation des ressources FPGA correspondent aux résultats réels. Les résultats, présentés dans la Table~\ref{tab:error_metrics}, confirment que les modèles mathématiques prédisent avec précision la consommation de LLUTs avec des valeurs \( R^2 \) élevées et de faibles erreurs.

\begin{table}[H]
    
    \centering
    \caption{Mesures d'erreur pour LLUT Models.}
    \label{tab:error_metrics}
    \begin{tabular}{l|cccc}
        \toprule
        \textbf{Bloc} & \textbf{EQM} & \textbf{EAM} & $\mathbf{R^2}$ & \textbf{EAMP (\%)} \\
        \hline
        \midrule
        $Conv_1$ & 16,244  & 3,054  & 0,997 & 3,038 \\
        $Conv_2$ & 0,498   & 0,538  & 0,941 & 2,134 \\
        $Conv_3$ & 0,00    & 0,00   & 1,00  & 0,00  \\
        $Conv_4$ & 0,379   & 0,518  & 0,989 & 1,342 \\
        \bottomrule
    \end{tabular}
\end{table}

\subsection{Prévision du pourcentage d'utilisation des ressources}

Pour valider l'efficacité des modèles mathématiques proposés, leur capacité à prédire la consommation de ressources dans différentes combinaisons de blocs de convolution est analysée. La Table~\ref{tab:prevue} montre les résultats pour les fonctions de convolution de précision 8 bits sur la plateforme Zynq UltraScale+ ZCU104, montrant comment les modèles peuvent aider à l'estimation des ressources et à la sélection des blocs de convolution pour optimiser l'utilisation du FPGA.

\begin{table}[h]
    \centering
    \caption{Consommation prévue des ressources (\%).}
    
    \label{tab:prevue}
    \begin{tabular}{p{0.55cm} p{0.55cm} p{0.55cm} p{0.75cm}|p{0.5cm}p{0.4cm}p{0.4cm}p{0.6cm}|p{0.5cm}}
        \toprule
        \multicolumn{4}{c|}{\textbf{\small{Nombre de blocs}}} & \multicolumn{4}{c|}{\textbf{\small{Usage des ressources (\%)}}} & \textbf{\small{Total}} \\
        \textbf{\small{$Conv_1$}} & \textbf{\small{$Conv_2$}} & \textbf{\small{$Conv_3$}} & \textbf{\small{$Conv_4$}} & \small{LLUT} & \small{FF}  & \small{DSP} & \small{CChain} & \textbf{\small{Conv.}} \\
        \hline
        1380 & 284  & 800  & 150  & 80,4 & 23,3 & 80,0 & 44,5 & 3564 \\
        1770 & 0    & 0    & 0    & 80,0 & 20,5 & 0,0  & 57,1 & 1770 \\
        0    & 1382 & 0    & 0    & 14,9 & 6,4  & 79,9  & 0,0 & 1382 \\
        0    & 0    & 1382 & 0    & 21,5 & 9,2  & 79,9  & 0,0 & 2764 \\
        0    & 0    & 0    & 691  & 11,1 & 3,3  & 79,9  & 0,0 & 1382 \\
        \bottomrule
    \end{tabular}
\end{table}

Dans la première ligne de la Table~\ref{tab:prevue}, les modèles ont été utilisés pour répartir stratégiquement les blocs, afin d’exploiter jusqu’à 80\% des ressources FPGA tout en augmentant le nombre de convolutions. Les lignes suivantes présentent la consommation attendue pour un seul type de bloc, permettant d’analyser l’impact individuel de chacun. Les résultats soulignent la capacité des modèles à soutenir des stratégies d’allocation des blocs, en fonction de la plateforme.

\section{Conclusion et Perspectives}
Cet article présente une bibliothèque de blocs de convolution conçue pour abstraire la complexité de la logique de contrôle et des opérations arithmétiques, facilitant ainsi l’implémentation de couches de convolution sur FPGA. Ces blocs sont paramétrables et adaptés à différentes contraintes de ressources, ce qui les rend particulièrement utiles dans des contextes où l’optimisation du matériel est essentielle. En outre, des modèles polynomiaux ont été développés pour estimer la consommation de ressources en fonction de la taille des données et des coefficients de la fonction de convolution, offrant ainsi un outil prédictif précieux dès les premières étapes de conception.

Les résultats expérimentaux ont révélé que les modèles sont efficaces pour optimiser l'allocation des blocs, en équilibrant l'utilisation de la logique, la disponibilité du DSP et les tailles des couches de convolution. En éliminant les itérations de synthèse répétées, la méthodologie accélère l'exploration de l'espace de conception, permettant l'évaluation de différentes configurations et l'identification des compromis optimaux entre la précision, le parallélisme et les contraintes de ressources. Les modèles proposés peuvent être adaptés à d’autres plateformes avec des architectures similaires, bien que l’étude repose sur un seul exemple.

Dans le cadre de travaux futurs, cette méthodologie pourrait être étendue à d’autres types de couches CNN et enrichie par l’intégration de critères supplémentaires tels que la consommation d’énergie ou la latence. Une telle extension permettrait d’optimiser plus finement les implémentations pour les systèmes embarqués, où les contraintes de performances et d’efficacité énergétique sont particulièrement critiques.

\vspace{5mm}
\noindent\textbf{Remerciements}

\noindent Ces travaux ont bénéficié d'une subvention publique financée par la région Auvergne-Rhône-Alpes, Grenoble Alpes Métropole et BPIFrance, dans le cadre du projet I-Démo Région "Green AI".

\bibliography{biblio.bib}


\end{document}